\documentclass[conference]{IEEEtran}
\IEEEoverridecommandlockouts
\usepackage{cite}
\usepackage{amsmath,amssymb,amsfonts}
\usepackage{algorithmic}
\usepackage{graphicx}
\usepackage{textcomp}
\usepackage{xcolor}
\usepackage{hyperref}
\usepackage{cuted}
\usepackage{amssymb}
\usepackage{booktabs}
\usepackage{float}
\usepackage{refcount}
\usepackage{cite}
\usepackage[T1]{fontenc}
\def\BibTeX{{\rm B\kern-.05em{\sc i\kern-.025em b}\kern-.08em
T\kern-.1667em\lower.7ex\hbox{E}\kern-.125emX}}
\begin{document}

\title{SongFormer: Scaling Music Structure Analysis with
Heterogeneous Supervision}

\author{
  \IEEEauthorblockN{Chunbo Hao$^{1,*}$, Ruibin Yuan$^{2,6,*}$, Jixun
    Yao$^{1}$, Qixin Deng$^{3,6}$, Xinyi Bai$^{4,6}$, Yanbo Wang$^{5}$,
    Wei Xue$^{2}$, Lei Xie$^{1,\dagger}$\thanks{$^{*}$Equal
  contribution. $^{\dagger}$Corresponding author.}}
  \IEEEauthorblockA{$^{1}$Audio, Speech and Language Processing Group
    (ASLP@NPU),\\
    School of Computer Science, Northwestern Polytechnical University\\
    $^{2}$Hong Kong University of Science and Technology \quad
    $^{3}$Northwestern University\\
    $^{4}$Cornell University \quad $^{5}$University of New South
    Wales \quad $^{6}$Multimodal Art Projection (M-A-P)\\
  cbhao@mail.nwpu.edu.cn, ryuanab@connect.ust.hk, lxie@nwpu.edu.cn}
}

\maketitle

\begin{abstract}

  Music structure analysis (MSA) underpins music understanding and
  controllable generation, yet progress has been limited by small,
  inconsistent corpora. We present SongFormer, a scalable framework
  that learns from heterogeneous supervision. SongFormer (i) fuses
  short- and long-window self-supervised learning representations to
  capture both fine-grained and long-range dependencies, and (ii)
  introduces a learned source embedding to enable training with
  partial, noisy, and schema-mismatched labels. To support scaling
  and fair evaluation, we release SongFormDB, the largest MSA corpus
  to date (over 14k songs spanning languages and genres), and
  SongFormBench, a 300-song expert-verified benchmark. On
  SongFormBench, SongFormer sets a new state of the art in strict
  boundary detection (HR.5F) and achieves the highest functional
  label accuracy, while remaining computationally efficient; it
  surpasses strong baselines and Gemini 2.5 Pro on these metrics and
  remains competitive under relaxed tolerance (HR3F). Code, datasets,
  and model are open-sourced at \url{https://github.com/ASLP-lab/SongFormer}.

\end{abstract}


\section{Introduction}
\label{sec:intro}

Music structure analysis (MSA)---segmenting a song into functionally
meaningful parts (e.g., \emph{intro}, \emph{verse}, \emph{chorus})
and detecting their boundaries---underpins music understanding and
controllable generation \cite{nieto2020audio,paulus2010state}. With
the rapid rise of music generation systems
\cite{ning2025diffrhythmblazinglyfastembarrassingly,yuan2025yuescalingopenfoundation,chen2025diffrhythmcontrollableflexiblefulllength,gong2025acestepstepmusicgeneration},
leveraging MSA as a structural prior has become increasingly
important. In practice, MSA is often cast as sequence labeling over
time \cite{wang2021supervised}.

However, current methods exhibit clear limitations that often result
in suboptimal performance and weak generalization. Public corpora
remain scarce and
heterogeneous~\cite{nieto2019harmonix,DBLP:conf/ismir/SmithBFRD11,DBLP:conf/ismir/GotoHNO02,mauch2009omras2}:
datasets are small, annotation schemes and formats vary considerably,
and access is frequently restricted. Consequently, much prior work
has been trained and evaluated on limited data, hindering
generalization~\cite{wang2022catchchorusverseintro}.
Methodologically, many systems are still trained from scratch rather
than leveraging pre-trained self-supervised (SSL) representations or
foundation audio models, and several systems rely on heavy
preprocessing such as beat tracking and source separation, which
increases complexity and further hinders scaling
\cite{taejun2023allinone,linkseg-hal-04665063}. While general-purpose
multimodal large language models (e.g.,
Gemini~2.5~Pro~\cite{DBLP:journals/corr/abs-2507-06261}) can produce
structure annotations, we observe that their temporal resolution is
too coarse for precise boundary detection and they may introduce
alignment or formatting issues in practice.

We present \textbf{SongFormer}, a simple, scalable framework that
leverages multi-resolution SSL representations and learns from
heterogeneous supervision. SongFormer fuses short- and long-window
SSL representations (30~s and 420~s) from
MuQ~\cite{zhu2025muqselfsupervisedmusicrepresentation} and
MusicFM~\cite{won2023foundationmodelmusicinformatics} to jointly
capture fine-grained and long-range dependencies. Additionally, a
learned source embedding is introduced to condition on dataset
provenance, enabling effective training with partial, noisy, and
schema-mismatched labels across diverse sources. To support
large-scale training and fair evaluation, we release
\textbf{SongFormDB}, a large corpus of over 14k songs spanning
languages and genres, and \textbf{SongFormBench}, a 300-song
expert-verified benchmark.

Experiments show that SongFormer achieves state-of-the-art strict
boundary detection (HR.5F) and functional label accuracy on
SongFormBench, outperforms all baselines on
RWC-Pop~\cite{DBLP:conf/ismir/GotoHNO02}, while efficiently
processing each song in just 2--4 s on an NVIDIA L40.

\begin{figure*}[htb]
  \centering
  \includegraphics
  {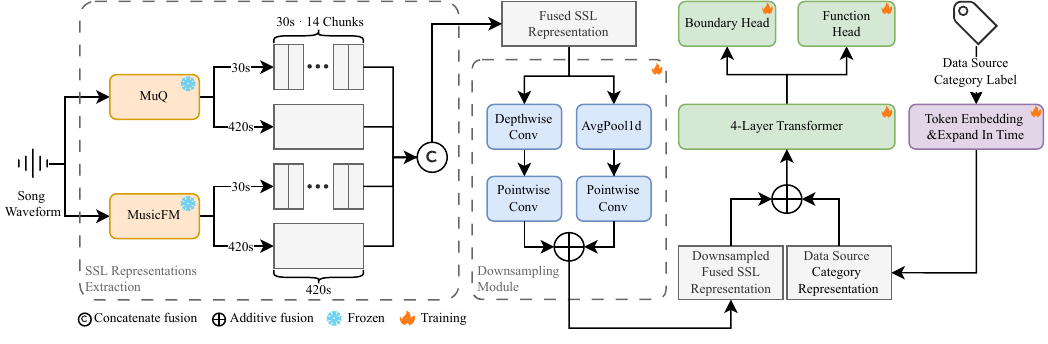}
  \caption{Overall architecture of our proposed SongFormer for music
  structure analysis.}
  \label{fig:SongFormer}
  \vspace{-1em}
\end{figure*}

\section{SongFormer}
\label{sec:method}

\subsection{Overview}

Fig.~\ref{fig:SongFormer} illustrates the architecture of SongFormer.
Multi-resolution SSL representations are extracted from the input
waveform, fused, and downsampled for computational efficiency. A
learned source embedding is added to enable heterogeneous
supervision. The combined representation is processed by a
Transformer encoder, followed by two task-specific heads that produce
boundary logits and functional logits.

During inference, we adopt the same post-processing as
All-In-One~\cite{taejun2023allinone}. Boundary timestamps are
identified by applying sigmoid activation, local maxima filtering,
and peak-picking to the boundary logits, partitioning the song into
segments. Each segment's functional label is then determined by the
class with the highest average frame-wise probability within that
segment from the softmaxed functional logits. The final output is an
ordered sequence of (start time, label) pairs:
\begin{equation}
  \{(t_0, l_0), (t_1, l_1), \dots, (t_{N-1}, l_{N-1}), (t_N, \texttt{end})\},
  \label{eq:res_sequence}
\end{equation}
where $t_i$ denotes the start time of each segment, $l_i \in
\mathcal{S}$ represents the functional label from the label set

\begin{equation}
  \mathcal{S} = \left\{\,
    \begin{aligned}
      &\texttt{intro}, \texttt{verse}, \texttt{pre-chorus}, \texttt{chorus}, \\
      &\texttt{bridge}, \texttt{inst}, \texttt{silence}, \texttt{outro}
    \end{aligned}
  \,\right\},
\end{equation}
and \texttt{end} marks the final segment's conclusion. Details of
each component are described in the following subsections.

\subsection{Fusion of Self-Supervised Learning Representations}

Previous work~\cite{taejun2023allinone} shows that training with
entire songs outperforms segment-level training, as a longer context
better captures structural dependencies. However, MuQ and MusicFM are
trained on 30~s windows, and extending inference windows degrades
performance due to context
mismatch~\cite{zhang2025temporaladaptationpretrainedfoundation}. To
leverage powerful representations while avoiding this degradation, we
extract SSL representations using two complementary temporal granularities.

\textbf{Local Representations.}
Given an input song waveform $\mathbf{x} \in \mathbb{R}^{T \cdot
f_s}$ with duration $T$ seconds and sampling rate $f_s$,
to capture fine-grained details, we segment the waveform into $N_1 =
\lceil T / \tau_l \rceil$ consecutive chunks of length $\tau_l =
30$~s. For each chunk, we extract local features as:
\begin{equation}
  \mathbf{H}_i^{\mathrm{L}} =
  \mathcal{E}\left(\mathbf{x}_{[(i-1)\tau_l \cdot f_s : \min(T \cdot
  f_s, i\tau_l \cdot f_s)]}\right), \; i = 1, \ldots, N_1,
\end{equation}
where $\mathcal{E}(\cdot)$ denotes a pre-trained audio encoder (MuQ
or MusicFM), $\mathbf{H}_i^{\mathrm{L}} \in \mathbb{R}^{L_{l,i}
\times D}$ is the local representation for the $i$-th chunk, and $D =
1024$ is the representation dimension.

\textbf{Global Representations.}
To capture overarching structural context, we process extended
windows of length $\tau_g = 420$\,s, sufficient to cover most songs
in their entirety. For longer pieces, we partition the input into
$N_2 = \lceil T / \tau_g \rceil$ windows and obtain
$\mathbf{H}_j^{\mathrm{G}} \in \mathbb{R}^{L_{g,j} \times D}$:
\begin{equation}
  \mathbf{H}_j^{\mathrm{G}} =
  \mathcal{E}\left(\mathbf{x}_{[(j-1)\tau_g \cdot f_s : \min(T \cdot
  f_s, j\tau_g \cdot f_s)]}\right), \; j = 1, \ldots, N_2.
\end{equation}

We employ two SSL encoders, MuQ and MusicFM, to extract
representations. For notational brevity, we denote MuQ as Q and
MusicFM as F in subsequent equations. The chunk-wise features are
concatenated along the temporal dimension to form complete sequence
representations:
\begin{align}
  \mathbf{H}^{\mathrm{L}}_{\mathrm{Q}} &=
  \left[\mathbf{H}^{\mathrm{L}}_{\mathrm{Q}, 1};  \cdots;
  \mathbf{H}^{\mathrm{L}}_{\mathrm{Q}, N_1}\right], \\
  \mathbf{H}^{\mathrm{G}}_{\mathrm{Q}} &=
  \left[\mathbf{H}^{\mathrm{G}}_{\mathrm{Q}, 1};  \cdots;
  \mathbf{H}^{\mathrm{G}}_{\mathrm{Q}, N_2}\right], \\
  \mathbf{H}^{\mathrm{L}}_{\mathrm{F}} &=
  \left[\mathbf{H}^{\mathrm{L}}_{\mathrm{F}, 1};  \cdots;
  \mathbf{H}^{\mathrm{L}}_{\mathrm{F}, N_1}\right], \\
  \mathbf{H}^{\mathrm{G}}_{\mathrm{F}} &=
  \left[\mathbf{H}^{\mathrm{G}}_{\mathrm{F}, 1};  \cdots;
  \mathbf{H}^{\mathrm{G}}_{\mathrm{F}, N_2}\right],
\end{align}
where $[\cdot\,;\cdot]$ denotes concatenation along the temporal
dimension. Each of the four representations has shape $\mathbb{R}^{L
\times D}$, where $L = \sum_{i=1}^{N_1} L_{l,i} = \sum_{j=1}^{N_2}
L_{g,j}$ is the total sequence length.

Subsequently, these four representations are fused via channel-wise
concatenation along the feature dimension, yielding a fused SSL
representation $\mathbf{H}^{\mathrm{fused}}\in \mathbb{R}^{L \times 4D}$:
\begin{equation}
  \mathbf{H}^{\mathrm{fused}} =
  \big[\mathbf{H}^{\mathrm{L}}_{\mathrm{Q}} \|
    \mathbf{H}^{\mathrm{G}}_{\mathrm{Q}}
    \| \mathbf{H}^{\mathrm{L}}_{\mathrm{F}} \|
  \mathbf{H}^{\mathrm{G}}_{\mathrm{F}}\big],
\end{equation}
where $[\cdot \| \cdot]$ denotes concatenation along the feature dimension.

To enhance computational efficiency, $\mathbf{H}^{\mathrm{fused}}$ is
processed by a downsampling module with two parallel branches:
depthwise separable convolution and average pooling with pointwise
convolution. Their outputs are fused via element-wise addition to
obtain $\mathbf{H}^{\mathrm{down}} \in \mathbb{R}^{\lfloor
\frac{L}{r} \rfloor \times D'}$, where $r=3$ is the downsampling
factor and $D' = 1024$ is the feature dimension. This reduces the
temporal resolution from 25~Hz to approximately 8.33~Hz while
preserving informative features.

\subsection{Heterogeneous Supervision
Strategies}\label{Heterogeneous_Supervision_Strategies}

The development of robust MSA models is constrained by the scarcity
of large-scale datasets and inconsistent annotations. To augment
training data, we incorporate external datasets with varying
annotation schemes and quality levels (details in
Section~\ref{sec:dataset}).
Following~\cite{wang2022catchchorusverseintro}, we map heterogeneous
label sets to a unified scheme while retaining \texttt{pre-chorus}
for its transitional role. However, this mapping remains imperfect
due to stylistic variation in labeling conventions and annotation
quality across datasets.

To address this challenge, we introduce a key component of our
heterogeneous supervision strategy: a learned source embedding.
Specifically, the source category $s \in \{\text{HX}, \text{E},
\text{H}, \text{G}\}$, corresponding to the four dataset categories
in Table~\ref{tab:dataset_statistics}, is mapped to a dense vector
$\mathbf{e}_s \in \mathbb{R}^{D'}$ via a learned embedding layer.
This embedding is then element-wise added to the downsampled representation:
\begin{equation}
  \tilde{\mathbf{H}}^{\mathrm{down}} = \mathbf{H}^{\mathrm{down}}
  \oplus \mathbf{e}_s,
\end{equation}
where $\tilde{\mathbf{H}}^{\mathrm{down}} \in \mathbb{R}^{\lfloor
\frac{L}{r} \rfloor \times D'}$ serves as the input to the subsequent
Transformer encoder. This conditioning mechanism enables the model to
learn source-specific annotation patterns and noise characteristics.
During inference, we set the source embedding to HarmonixSet ($s =
\text{HX}$), which has complete eight-category annotations. Using
other source embeddings would cause label mismatches. This ensures
label consistency while benefiting from diverse training data to
improve generalization.

\subsection{Training Objective}\label{training_objective}

The downsampled representation $\tilde{\mathbf{H}}^{\mathrm{down}}$
is fed into a 4-layer Transformer encoder to capture long-range
temporal dependencies:
\begin{equation}
  \mathbf{H}^{\mathrm{enc}} =
  \mathrm{Transformer}\left(\tilde{\mathbf{H}}^{\mathrm{down}}\right)
  \in \mathbb{R}^{\lfloor \frac{L}{r} \rfloor \times D''},
\end{equation}
where $D'' = 512$ is the hidden dimension of the Transformer encoder.
$\mathbf{H}^{\mathrm{enc}}$ is then passed through two task-specific
prediction heads to obtain frame-level activation curves: boundary
probabilities $\mathbf{p}^{\mathrm{bdy}} \in \mathbb{R}^{\lfloor
\frac{L}{r} \rfloor}$ and function probabilities
$\mathbf{P}^{\mathrm{func}} \in \mathbb{R}^{\lfloor \frac{L}{r}
\rfloor \times C}$, where $C = 8$ is the number of structural labels.

For boundary detection, we employ a binary cross-entropy loss
$\mathcal{L}_{\mathrm{BCE}}$. Additionally, we introduce a
boundary-aware one-dimensional (1D) Total Variation (TV)
loss~\cite{rudin1992nonlinear} to encourage temporal smoothness by
penalizing abrupt fluctuations at nonboundary positions.
Given the boundary predictions $\mathbf{p}^{\mathrm{bdy}}$, we
compute the frame-to-frame differences as:
\begin{equation}
  \Delta \mathbf{p}^{\mathrm{bdy}}_{t} =
  \mathbf{p}^{\mathrm{bdy}}_{t+1} - \mathbf{p}^{\mathrm{bdy}}_{t}.
\end{equation}

We define the basic TV penalty as $|\Delta
\mathbf{p}^{\mathrm{bdy}}_{t}|^\beta$ with $\beta = 0.6$. The penalty
weight $\alpha$ is set to $0.1$ if $t$ lies within a boundary region
(i.e., where the smoothed ground-truth boundary activation exceeds
$0.01$), and $1$ otherwise. The resulting loss is formulated as:
\begin{equation}
  \mathcal{L}_{\text{TV}} =
  \frac{1}{\lfloor \frac{L}{r} \rfloor-1}
  \sum_{t=1}^{\lfloor \frac{L}{r} \rfloor-1} \alpha_t \, |\Delta
  p^{bdy}_{t}|^\beta.
\end{equation}

For functional prediction, we employ a frame-wise cross-entropy loss
$\mathcal{L}_{\mathrm{CE}}$ and softmax focal loss
$\mathcal{L}_{\mathrm{Focal}}$~\cite{lin2017focal} to focus on uncertain frames.

Finally, the total loss is defined as the weighted sum of boundary
and function objectives, with $\lambda = 0.2$, $\lambda_{\text{TV}} =
0.05$, and $\lambda_{\text{Focal}} = 0.2$:
\begin{equation}
  \begin{split}
    \mathcal{L} = \;&\lambda \left( \mathcal{L}_{\mathrm{BCE}} +
    \lambda_{\mathrm{TV}} \, \mathcal{L}_{\mathrm{TV}} \right) \\
    &+ (1-\lambda) \left( \mathcal{L}_{\mathrm{CE}} +
    \lambda_{\mathrm{Focal}} \, \mathcal{L}_{\mathrm{Focal}} \right).
  \end{split}
\end{equation}

To further handle the heterogeneity across datasets, we introduce
task-specific masks at the frame level for boundary and functional
losses. In SongFormDB-Hook, the function loss is computed only within
valid segments, with the temporal window for boundary loss extended
by 5~s on each side. In SongFormDB-Gem, only the functional loss is
optimized, thereby mitigating the impact of inaccurate segment
boundaries on model training.

\section{Dataset}

\label{sec:dataset}

\begin{table}[t]
  \centering

  \caption{Dataset statistics. The upper and lower sections show
    SongFormDB and SongFormBench, respectively. In SongFormDB-Hook,
  partial segments of each song are annotated.}
  \vspace{-5pt}
  \label{tab:dataset_statistics}
  \begin{tabular}{p{3.4cm}c|ccc}
    \toprule
    \textbf{Dataset} &  \textbf{Abbr.}&\textbf{Train} & \textbf{Eval}
    & \textbf{Test} \\
    \midrule
    SongFormDB-HX&  HX&512   & 200 & --\\
    SongFormDB-Ext&  E&4,314 &  -- & --  \\
    SongFormDB-Hook&  H&5,933 &  -- & --  \\
    SongFormDB-Gem& G& 4,387& --&--\\
    \midrule
    SongFormBench-HarmonixSet& BHX& --& --&200\\
    SongFormBench-CN& BC& --& --&100\\

    \bottomrule
  \end{tabular}
  \vspace{-10pt}
\end{table}

We establish SongFormDB, a large-scale collection of annotated songs
for training, and SongFormBench, a high-quality benchmark for
evaluation, to address data limitations.
The mapping rules from~\cite{wang2022catchchorusverseintro} are
adopted while preserving the \texttt{pre-chorus} label to better
capture transitional passages that build tension and anticipation
before the chorus, resulting in a unified label set of 8 categories.
To facilitate reproducibility, we will publicly release structural
annotations, pre-extracted Mel spectrograms, and metadata for both
SongFormDB and SongFormBench. The following subsections present
detailed descriptions of these datasets.

\subsection{SongFormDB}

For SongFormDB-HX, we select 512 songs from the official
HarmonixSet~\cite{nieto2019harmonix} for training and 200 for
validation. We employ rule-based corrections to refine the
annotations, such as replacing erroneous ``outro'' labels at the
beginning of songs with ``intro'' and calibrating silence segment durations.

Songs in the remaining subsets of SongFormDB are sourced from YouTube.
For SongFormDB-Ext, structural labels with timestamps are derived
from text-audio alignment metadata. We employ the Singing-Oriented
Forced Aligner (SOFA)\footnote{SOFA:
\url{https://github.com/qiuqiao/SOFA}} to detect misalignments:
segments whose predicted start times deviate from the original
timestamps by more than 1 second are marked as errors, and data with
error rates exceeding 10\% are discarded. This yields 4,314 well-aligned songs.

For SongFormDB-Hook, we collect 5,933 songs with accurate structural
annotations covering partial segments. These segments may be
contiguous or noncontiguous, and each is labeled with either a single
structural label or multiple functionally similar labels---all
considered correct---with one randomly selected during training.

For SongFormDB-Gem, we sample songs across 47 languages, balance
tempo distribution using 10 bins, and ensure broad genre coverage. We
then leverage Gemini 2.5 Pro Preview
(05-06)~\cite{DBLP:journals/corr/abs-2507-06261} to generate
structural annotations, providing a system prompt, the audio, and
format instructions as input. The model outputs undergo rigorous
filtering to remove malformed formats, nonmonotonic segment
timestamps, and abnormal total durations, ultimately yielding 4,387 annotations.
Careful empirical analysis reveals that Gemini's boundary predictions
exhibit coarse temporal resolution (up to 2\,s error), while its
function labels are generally accurate. Since manual annotation is
both expensive and difficult to obtain at scale, we leverage only
these accurate function labels by employing the learned source
embedding (\ref{Heterogeneous_Supervision_Strategies}) and disabling
the boundary loss for SongFormDB-Gem (\ref{training_objective}) to
handle its distinct characteristics, thereby improving SongFormer's
generalization.

\subsection{SongFormBench}

To establish a standardized benchmark for fair comparison in future
MSA research, we introduce SongFormBench, a high-quality benchmark
comprising 200 songs from HarmonixSet and 100 Chinese songs. All
annotations are meticulously revised by expert annotators through a
rigorous cross-checking process based on the original audio and
dataset labels, with publicly available metadata platforms (e.g.,
  MusixMatch\footnote{MusixMatch: \url{https://www.musixmatch.com}},
Genius\footnote{Genius: \url{https://genius.com}}) consulted for
reference. Audio fingerprint comparison is conducted to ensure no
overlap between SongFormBench and the training set.

\section{Experiments}

\label{sec:exp}

\subsection{Evaluation Metrics}

Following the widely adopted evaluation protocol in MSA literature,
we use the following metrics: \textbf{(i) HR.5F} and \textbf{(ii)
HR3F}: The F-measure of boundary hit rate within 0.5 and 3 seconds,
respectively. These metrics assess segmentation accuracy and penalize
both over-segmentation and under-segmentation errors. \textbf{(iii)
Accuracy (ACC)}: As MSA models partition each song into
non-overlapping, exhaustive segments as defined
in~\eqref{eq:res_sequence}, ACC measures the proportion of frames
whose assigned label from the model's final output matches the ground
truth, reflecting the accuracy of functional label predictions.

\subsection{Experimental Settings}

As shown in Table~\ref{tab:dataset_statistics}, SongFormDB is used
for training, while SongFormBench serves as the test set. During
preprocessing, songs exceeding 420 s are split into 420 s segments;
shorter ones retain their original length.

We adopt the official open-source implementations for both
pre-trained SSL models, using the default pre-trained weights for MuQ
and the MusicFM-MSD variant for MusicFM.
Segment boundaries are smoothed with a Gaussian kernel spanning 10
frames on each side (approximately 2.4~s total at 8.33~Hz).
SongFormer employs a 4-layer Transformer with a hidden size of 512.
Training uses a batch size of 8 and a cosine learning rate schedule
(peak $1 \times 10^{-4}$, with 300 warm-up steps, and decayed to zero
after 12,000 steps). Early stopping is triggered if HR.5F or ACC does
not improve for three consecutive validations on the SongFormDB-HX
validation set. Experiments are repeated with three random seeds on a
single NVIDIA L40 GPU; averaged results are reported.

\subsection{Evaluation Protocol}

For fair comparison with prior work, we standardize the outputs of
all models to conform to the (start time, label) pair format
in~\eqref{eq:res_sequence}, which partitions each song into
non-overlapping, exhaustive segments. In both the main experiments
and ablation studies, we follow the mainstream seven-category
evaluation scheme from~\cite{wang2022catchchorusverseintro} using the
same evaluation script.
Specifically, although SongFormer outputs eight functional
categories---retaining \texttt{pre-chorus} to better capture
transitional passages that build tension before the chorus---we are
constrained to consolidate \texttt{pre-chorus} into \texttt{verse}
during evaluation. This compromise is necessitated by two factors.
First, prior
methods~\cite{wang2022catchchorusverseintro,taejun2023allinone,zhu2025muqselfsupervisedmusicrepresentation,linkseg-hal-04665063}
all report results exclusively under the seven-category scheme,
leaving no eight-category baseline for comparison. Second, several
earlier models do not release public inference code, precluding
re-evaluation under a finer-grained scheme.

We evaluate on both SongFormBench and
RWC-Pop~\cite{DBLP:conf/ismir/GotoHNO02}, the latter comprising 100
annotated songs from the popular music subset of the RWC dataset. For
models with publicly available inference code, we standardize outputs
as follows: All-In-One maps \texttt{start}/\texttt{end} to
\texttt{silence} and \texttt{break}/\texttt{solo} to \texttt{inst},
normalizing to seven categories; LinkSeg is inferred using its
seven-category checkpoint; Gemini 2.5 Pro Preview (05-06) is prompted
with a system prompt, the audio, and expected output format to
generate structured annotations.

For models that are difficult to reproduce for inference, we resort
to citing results from prior work. Specifically,
Harmonic-CNN~\cite{DBLP:conf/icassp/WonCNS20} and
SpecTNT~\cite{lu2021spectnttimefrequencytransformermusic}, originally
proposed for general music information retrieval (MIR) tasks, are
adapted for MSA in~\cite{wang2022catchchorusverseintro}; we report
reproduced results from this work. For
MusicFM~\cite{won2023foundationmodelmusicinformatics}, we use results
from~\cite{zhang2025temporaladaptationpretrainedfoundation}, while
results for other models are taken from their original papers.

\subsection{Main Results}

As shown in Table~\ref{tab:SongFormBench_results}, SongFormer
achieves state-of-the-art performance on ACC and HR.5F across
SongFormBench and RWC-Pop, while remaining competitive on HR3F.

\begin{table}[tb]
  \centering

  \caption{Model performance on SongFormBench and
    RWC-Pop~\cite{DBLP:conf/ismir/GotoHNO02}. $\star$ denotes cited
    results from prior works with closed-source models. (HX, E, H, G)
  refer to training datasets in Table~\ref{tab:dataset_statistics}.}
  \vspace{-5pt}
  \begin{tabular}{l|lll}
    \toprule
    \textbf{Method} & \textbf{ACC} & \textbf{HR.5F} & \textbf{HR3F} \\
    \midrule
    \multicolumn{4}{c}{\textbf{SongFormBench-HarmonixSet}}\\
    \midrule
    Harmonic-CNN~\cite{wang2022catchchorusverseintro} $\star$& 0.680
    & 0.559 & --\\
    SpecTNT (24~s, CTL)~\cite{wang2022catchchorusverseintro} $\star$&
    0.701 & 0.570 & --\\
    SpecTNT (36~s, CTL)~\cite{wang2022catchchorusverseintro} $\star$&
    0.723 & 0.558 & --\\
    All-In-One~\cite{taejun2023allinone} & 0.740& 0.596& 0.730\\
    MusicFM-Zhang et
    al.~\cite{zhang2025temporaladaptationpretrainedfoundation}
    $\star$& 0.725& 0.640&0.729\\
    $\text{MuQ}_{\text{iter}}$~\cite{zhu2025muqselfsupervisedmusicrepresentation}
    $\star$& 0.772& --&--\\
    LinkSeg-7Labels~\cite{linkseg-hal-04665063} & 0.780& 0.630& 0.762\\
    TA (Zhang et al.,
    2025)~\cite{zhang2025temporaladaptationpretrainedfoundation}
    $\star$& 0.787& 0.610&\underline{0.801}\\
    Gemini 2.5 Pro~\cite{DBLP:journals/corr/abs-2507-06261}& 0.748&
    0.423&\textbf{0.813}\\
    \midrule
    SongFormer (HX)& 0.795& \textbf{0.703}&0.784\\
    SongFormer (HX+E+H)& \underline{0.806}& \underline{0.697}&0.780\\
    SongFormer (HX+E+H+G)& \textbf{0.807}& 0.696&0.780\\
    \midrule
    \multicolumn{4}{c}{\textbf{SongFormBench-CN}}\\
    \midrule
    All-In-One~\cite{taejun2023allinone} & 0.834& 0.563& 0.771\\
    LinkSeg-7Labels~\cite{linkseg-hal-04665063} & 0.828& 0.518&0.757\\
    Gemini 2.5 Pro\cite{DBLP:journals/corr/abs-2507-06261}& 0.806& 0.412&0.833\\
    \midrule
    SongFormer (HX)& 0.848& 0.675& \textbf{0.856}\\
    SongFormer (HX+E+H)& \underline{0.890}& \textbf{0.690}& \underline{0.852}\\
    SongFormer (HX+E+H+G)& \textbf{0.891}& \underline{0.688}& 0.851\\
    \midrule
    \multicolumn{4}{c}{\textbf{RWC-Pop}}\\
    \midrule
    Harmonic-CNN~\cite{wang2022catchchorusverseintro} $\star$& 0.646&
    0.571& --\\
    SpecTNT (24~s, CTL)~\cite{wang2022catchchorusverseintro} $\star$&
    0.675& 0.623& --\\
    MusicFM-Zhang et
    al.~\cite{zhang2025temporaladaptationpretrainedfoundation}
    $\star$& 0.680& 0.636& 0.764\\
    LinkSeg~\cite{linkseg-hal-04665063}& 0.747& 0.648& 0.786\\
    TA (Zhang et al.,
    2025)~\cite{zhang2025temporaladaptationpretrainedfoundation}
    $\star$& 0.779& 0.506& 0.691\\
    \midrule
    SongFormer (HX)& 0.787& \textbf{0.651}& 0.795\\
    SongFormer (HX+E+H)& \textbf{0.814}& \underline{0.650}& \textbf{0.804}\\
    SongFormer (HX+E+H+G)& \underline{0.812}& \underline{0.650}&
    \underline{0.800}\\
    \bottomrule
  \end{tabular}

  \label{tab:SongFormBench_results}
  \vspace{-15pt}
\end{table}

On SongFormBench-HarmonixSet, SongFormer (HX+E+ H+G) achieves the
highest ACC of 0.807, surpassing TA (0.787) and LinkSeg-7Labels
(0.780). For strict boundary accuracy (HR.5F), SongFormer (HX)
attains 0.703, substantially outperforming LinkSeg (0.630) and
All-In-One (0.596). While Gemini 2.5 Pro achieves a slightly higher
relaxed boundary score (HR3F = 0.813), its strict boundary accuracy
(0.423) is notably inferior, indicating imprecise segment boundaries.

On SongFormBench-CN, SongFormer's advantages are more pronounced.
SongFormer (HX+E+H+G) achieves the highest ACC of 0.891, while
SongFormer (HX+E+H) attains the best HR.5F of 0.690—significantly
exceeding All-In-One (0.563) and LinkSeg (0.518). SongFormer (HX)
also achieves the highest HR3F of 0.856, surpassing Gemini 2.5 Pro (0.833).

To ensure unbiased evaluation, we also report results on
RWC-Pop~\cite{DBLP:conf/ismir/GotoHNO02}, a widely used dataset
entirely excluded from our training data. On this external benchmark,
all SongFormer variants achieve state-of-the-art performance across
all metrics, demonstrating that our improvements generalize beyond
SongFormBench.

Overall, these results demonstrate that: \textbf{(i)} SongFormer
achieves superior label accuracy and sharper boundary predictions,
establishing it as a more precise and generalizable framework for
MSA; \textbf{(ii)} SongFormer (HX), trained solely on 512
human-annotated songs in HarmonixSet without any Gemini-generated
labels, already surpasses Gemini 2.5 Pro and all other baselines on
both ACC and HR.5F across all test sets, ruling out potential teacher
bias from SongFormDB-Gem and confirming the effectiveness of our
architectural design rather than mere reliance on data scale;
\textbf{(iii)} scaling from HX to HX+E+H+G further improves ACC while
slightly reducing HR.5F due to timestamp inaccuracies---precisely the
issue that the learned source embedding is designed to mitigate.
\textbf{(iv)} although we set the source embedding to HarmonixSet
during inference, SongFormer achieves state-of-the-art results on
RWC-Pop and SongFormBench-CN, demonstrating that the HX embedding
does not merely inject a ``HarmonixSet-style'' prior to inflate
performance on the HarmonixSet-derived test set, but enables genuine
cross-dataset generalization.

Moreover, thanks to its efficient design and avoidance of complex
preprocessing, SongFormer achieves the fastest inference speed
(Table~\ref{tab:infer-time}), demonstrating strong practical
potential for real-world applications.

\begin{table}[tb]
  \centering
  \caption{Average end-to-end inference time per song on
    SongFormBench-HarmonixSet, measured on a single NVIDIA L40 GPU,
    covering the entire pipeline from raw waveform input to final
  structured output.}
  \vspace{-5pt}
  \label{tab:infer-time}
  \begin{tabular}{l|c}
    \toprule
    \textbf{Method} & \textbf{Time (s)} \\
    \midrule
    LinkSeg-7Labels & 3--5 \\
    All-In-One & 9--12 \\
    Gemini 2.5 Pro & 30--90 \\
    \midrule
    \textbf{SongFormer} & \textbf{2--4} \\
    \bottomrule
  \end{tabular}
\end{table}

\subsection{Ablation Study}

\begin{figure}[tb]
  \centering

  \includegraphics
  {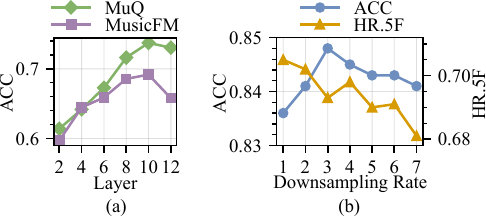}
  \vspace{-5pt}
  \caption{Effect of representation layers and downsampling on model
    performance. (a) ACC of MuQ and MusicFM across layers. (b) ACC and
  HR.5F under different downsampling rates.}
  \label{fig:ablation_fig}
  \vspace{-15pt}
\end{figure}

In Fig.~\ref{fig:ablation_fig}(a), we train simple multi-layer
perceptrons (MLPs) with task-specific heads on features from
individual layers of each SSL encoder using the HX dataset. Both MuQ
and MusicFM achieve peak performance at the 10th layer, which we
adopt for all subsequent experiments.
Fig.~\ref{fig:ablation_fig}(b) examines the effect of temporal
resolution by varying the hop size in the downsampling module, with
architecture otherwise identical to the main model. As the
downsampling factor increases, hit rate (HR) consistently declines
due to reduced resolution, while ACC initially rises before
declining. A downsampling factor of 3 offers the best trade-off
between computational efficiency and prediction accuracy.

Both Table~\ref{tab:Step-by-step} and
Table~\ref{tab:multi-resolution-self-supervised} present systematic
ablation studies, with all experiments conducted under identical
settings: models are trained on the HX, E, and H datasets and
evaluated on the complete SongFormBench comprising 300 songs.

Table~\ref{tab:Step-by-step} presents a systematic ablation study on
the contribution of each component in SongFormer. The results show
that multi-resolution representations, downsampling, and learned
source embedding substantially improve performance, while combining
MuQ and MusicFM yields more robust representations. Removing any of
these components degrades performance, underscoring their importance.

Notably, learned source embedding plays a crucial role in handling
our heterogeneous training data, which may contain annotation noise
or quality issues. Without this embedding, large-scale but
lower-quality annotations can negatively impact the model's ability
to learn from precisely labeled data, such as HarmonixSet. By
incorporating learned source embedding, the model distinguishes
between different data sources during training. At inference time, we
specify the Harmonixset embedding to leverage its precise annotations
while still benefiting from the generalization learned across
heterogeneous sources.

\begin{table}[tp]
  \centering

  \caption{Impact of different components on model performance.
    Q: MuQ; F: MusicFM; M: multi-resolution SSL representations; D:
    downsampling strategy; B: backend architecture (T: Transformer,
    M: linear layer); S: learned source embedding.
  }
  \vspace{-5pt}
  \label{tab:Step-by-step}
  \begin{tabular}{cccccc|ccc}
    \toprule
    \textbf{Q}& \textbf{F}& \textbf{M} & \textbf{D} & \textbf{B} &
    \textbf{S} & \textbf{ACC} & \textbf{HR.5F} & \textbf{HR3F} \\
    \midrule
    $\checkmark$ & $\checkmark$ & $\checkmark$ & $\checkmark$ & T &
    $\checkmark$ & \textbf{0.848}& \textbf{0.693}& \textbf{0.816}\\
    \midrule
    $\checkmark$ & $\checkmark$ & $\checkmark$ & $\checkmark$ & T &
    -- & \underline{0.825} & 0.685 & 0.801 \\
    $\checkmark$ & $\checkmark$ & $\checkmark$ & $\checkmark$ & M& --
    & 0.797& 0.688& \underline{0.803}\\
    $\checkmark$ & $\checkmark$ & $\checkmark$ & -- & M& -- & 0.789&
    \underline{0.690}& 0.802\\
    $\checkmark$ & $\checkmark$ & -- & -- & M& -- & 0.754& 0.688& 0.802\\
    $\checkmark$ & -- & -- & -- & M& -- & 0.749& 0.686& 0.802\\
    -- & $\checkmark$ & -- & -- & M& -- & 0.718& 0.669& 0.786\\
    \bottomrule
  \end{tabular}
  \vspace{-5pt}
\end{table}

In Table~\ref{tab:multi-resolution-self-supervised}, we evaluate the
impact of multi-resolution SSL representations. Using 30~s SSL
embeddings for 30~s SongFormer (\textit{No.~2}) yields the lowest
ACC, as the short window fails to capture full-song context.
Extending the SSL window to 420~s (\textit{No.~2} → \textit{No.~3})
improves ACC but lowers HR, reflecting a mismatch between the 420~s
embeddings and the SSL model’s training window, consistent
with~\cite{zhang2025temporaladaptationpretrainedfoundation}. In
contrast, concatenating 30~s embeddings into a 420~s input
(\textit{No.~4}) provides substantial gains, aligning SSL inference
with its training window while enabling longer sequence modeling.
Combining this with 420~s embeddings (\textit{No.~1}) achieves the
best performance, underscoring the advantage of multi-resolution SSL
representations.

\begin{table}[tb]
  \centering

  \caption{Model performance with different SSL embeddings (MuQ and
    MusicFM) and time window configurations. 30~s and 420~s indicate
    the use of 30~s or 420~s SSL embeddings, while Duration refers to
  the input duration for SongFormer.}
  \vspace{-5pt}
  \label{tab:multi-resolution-self-supervised}
  \begin{tabular}{c|ccc|ccc}
    \toprule
    \textbf{No.}&\textbf{30~s} & \textbf{420~s} & \textbf{Duration} &
    \textbf{ACC} & \textbf{HR.5F} & \textbf{HR3F} \\
    \midrule
    1&\checkmark &      \checkmark & 420~s& \textbf{0.848} &
    \textbf{0.693} & \underline{0.816} \\
    \midrule
    2&\checkmark &      & 30~s& 0.782 & \underline{0.689} & \textbf{0.817} \\
    3&& \checkmark & 420~s& 0.834 & 0.677 & 0.802 \\

    4& \checkmark & & 420~s& \underline{0.835} & \textbf{0.693} &0.812 \\
    \bottomrule
  \end{tabular}
  \vspace{-15pt}
\end{table}

\section{Conclusion}

SongFormer is a scalable and efficient framework for MSA that fuses
multi-resolution SSL representations with heterogeneous supervision.
Extensive experiments and ablations confirm robust generalization and
validate each component. To mitigate data scarcity, we release
SongFormDB—the largest training corpus to date—and SongFormBench, a
curated benchmark (200 manually revised HarmonixSet, 100 Chinese
songs), enabling high-quality, fair, and reproducible evaluation and
advancing the integration of MSA into music information retrieval and
controllable music generation.

\bibliographystyle{IEEEbib}
\bibliography{icme2026references}

\end{document}